\documentclass{jnmp}

\usepackage{amsfonts,amssymb}

\usepackage{amssymb}
\usepackage{eucal}

\def\openone{\leavevmode\hbox{\small1\kern-3.3pt\normalsize1}}

\def\bbbz{\Bbb{Z}}

\def\bbbc{{\Bbb C}}

\def\bbbz{{\Bbb Z}}
\def\openone{\leavevmode\hbox{\small1\kern-3.3pt\normalsize1}}

\usepackage{epsf}

\def\newpic#1{%
   \def\emline##1##2##3##4##5##6{%
      \put(##1,##2){\special{em:point #1##3}}%
      \put(##4,##5){\special{em:point #1##6}}%
      \special{em:line #1##3,#1##6}}}

\newpic{}

\def\bbbc{{\Bbb C}}

\def\bbbz{{\Bbb Z}}

\def\openone{\leavevmode\hbox{\small1\kern-3.3pt\normalsize1}}

\begin{document}

\renewcommand{\evenhead}{V. A. Atanasov, V. S. Gerdjikov, G. G. Grahovski and N. A.
Kostov}
\renewcommand{\oddhead}{Fordy-Kulish models and  spinor Bose-Einstein condensates}

\thispagestyle{empty}

\FirstPageHead{*}{*}{20**}{\pageref{firstpage}--
\pageref{lastpage}}{Article}

\label{firstpage}

\Name{Fordy-Kulish model and  spinor Bose-Einstein condensate}

\Author{V. A. Atanasov$^{1,2}$ , V. S. Gerdjikov$^1$ , G. G. Grahovski$^{1,3}$ , N. A.
Kostov$^1$ }

\Address{$^1$ Institute for Nuclear Research and Nuclear Energy,
Bulgarian Academy of Sciences,\\ 72 Tsarigradsko chauss\'ee,
1784 Sofia, Bulgaria\\
$^3$ Laboratoire de Physique Th\'eorique et
Mod\'elisation, Universit\'e de Cergy-Pontoise,\\ 2 avenue A.
Chauvin, F-95302 Cergy-Pontoise Cedex,
France\\
$^3$ School of
Electronic Engineering, Dublin City University, Glasnevin, Dublin
9, Ireland\\
~~E-mail: victor@inrne.bas.bg, \quad gerjikov@inrne.bas.bg, \quad grah@inrne.bas.bg, \quad nakostov@inrne.bas.bg}

\Date{Received Month *, 200*; Accepted Month *, 200*}

\begin{abstract}
A three-component nonlinear Schrodinger-type model which describes
spinor Bose-Einstein condensate ( BEC) is considered. This model
is integrable by the inverse scattering method and using
Zakha\-rov-Shabat dressing method we obtain three types of soliton
solutions. The multi-component nonlinear Schr\"{o}dinger type
models related to symmetric spaces ${\bf C.I}\simeq {\rm
Sp(4)}/{\rm U(2)}$ is studied.
\end{abstract}

\section{Introduction}

The dynamics of spinor BEC is described by a three-component
Gross-Pitaevskii (GP) system of equations. In the one-dimensional
approximation  the GP system goes into the following nonlinear
Schr\"{o}dinger (MNLS) equation in (1D) $x$-space \cite{IMW04}:
\begin{eqnarray}\label{eq:1}
&& i\partial_{t} \Phi_{1}+\partial^{2}_{x} \Phi_{1}+2(|\Phi_{1}|^2
+2|\Phi_{0}|^2) \Phi_{1} +2\Phi_{-1}^{*}\Phi_{0}^2=0, \nonumber \\
&& i\partial_{t} \Phi_{0}+\partial^{2}_{x}
\Phi_{0}+2(|\Phi_{-1}|^2
+|\Phi_{0}|^2+|\Phi_{1}|^2) \Phi_{0} +2\Phi_{0}^{*}\Phi_{1}\Phi_{-1}=0,\\
&& i\partial_{t}\Phi_{-1}+\partial^{2}_{x}
\Phi_{-1}+2(|\Phi_{-1}|^2+ 2|\Phi_{0}|^2) \Phi_{-1}
+2\Phi_{1}^{*}\Phi_{0}^2=0. \nonumber
\end{eqnarray}
We consider BEC of alcali atoms in the $F=1$ hyperfine state,
elongated in $x$ direction and confined in the transverse
directions $y,z$ by purely optical means. Thus the assembly of
atoms in the $F=1$ hyperfine state can be described by a
normalized spinor wave vector
\begin{eqnarray} {\bf\Phi}(x,t)=(\Phi_1(x,t), \Phi_0(x,t),
\Phi_{-1}(x,t))^{T}\end{eqnarray} whose components are labelled by
the values of $m_F =1,0,-1$. The above model is integrable by
means of inverse scattering transform method \cite{IMW04}. It also
allows an exact description of the dynamics and interaction of
bright solitons with spin degrees of freedom. Matter-wave solitons
are expected to be useful in atom laser, atom interferometry and
coherent atom transport. It could contribute to the realization of
quantum information processing or computation, as a part of new
field of atom optics. Lax pairs and geometric interpretation of
the model (\ref{eq:1}) are given in \cite{ForKu*83}. Darboux
transformation for this special integrable model is developed in
\cite{LLMML05}. The aim of present paper is to show that the
system (\ref{eq:1}) is related to symmetric space ${\bf C.I}\simeq
{\rm Sp(4)}/{\rm U(2)}$ (in the Cartan classification \cite{Helg})
with canonical $\bbbz_2$-reduction and has natural Lie algebraic
interpretation. The model allows also a special class of soliton
solutions. We will show that they can be obtained by a suitable
modification of the generalization of the so-called ``dressing
method'', proposed in \cite{Za*Mi}.


\section{Solving the model: algebraic and analytic aspects.}\label{ssec:2.3a}

The model (\ref{eq:1}) belongs to the class of multicomponent NLS
equations that can be solved by the inverse scattering method
\cite{ZMNP,FaTa}. It is a particular case of the MNLS related to
the  ${\bf C.I}$ type symmetric space ${\rm Sp(4)}/{\rm U(2)}$
\cite{ForKu*83}. These MNLS systems allow Lax representation with
the generalized Zakharov--Shabat system as the Lax operator:
\begin{eqnarray} \label{eq:Lax-MNLS}
L\psi (x,t,\lambda ) \equiv i {d\psi \over d x} + (Q(x,t) -
\lambda J)\psi (x,t,\lambda )=0.
\end{eqnarray}
where $J $ and $Q(x,t)$ are $4\times 4$ matrices: $ J =
\mbox{diag}\, (1,1,-1,-1)$ and $Q(x,t)$ is a block-off-diagonal
matrix:
\begin{eqnarray}\label{eq:C2.1}
&& Q(x,t)=\left( \begin{array}{cc}  0  &  {\bf q}(x,t) \\
  {\bf p}(x,t)  &  0 \\  \end{array} \right), \quad
 {\bf q}(x,t)=\left( \begin{array}{cccc}
 \Phi_{0}(x,t) & -\Phi_{1}(x,t) \\ \Phi_{-1}(x,t) &  -\Phi_{0}(x,t)
  \end{array} \right),\nonumber \\&& {\bf p}(x,t)=\left(
\begin{array}{cccc} \Phi_{0}^*(x,t) & \Phi_{-1}^*(x,t) \\
 -\Phi_{1}^*(x,t) &  -\Phi_{0}^*(x,t) \\ \end{array} \right).
\end{eqnarray}
Solving the direct and the inverse scattering problem for $L$ uses
the Jost solutions $\phi=(\phi^{+},\phi^{-})$ and
$\psi=(\psi^{-},\psi^{+})$ of (\ref{eq:Lax-MNLS}) which are
defined by, see \cite{VSG2} and the references therein:
\begin{equation}
 \lim_{x \to -\infty} \phi(x,t,\lambda) e^{  i \lambda J x
 }=\openone,\qquad
  \lim_{x \to \infty}\psi(x,t,\lambda) e^{  i \lambda J x
 } = \openone
 \end{equation}
These definitions are compatible with the class of smooth
potentials $Q(x,t)$  vanishing sufficiently rapidly at $x \to \pm
\infty$. It can be shown that $\phi^{+}$ and $\psi^{+}$ (resp.
$\phi^{-}$ and $\psi^{-}$) composed by $4$ rows and $2$ columns
are analytic in the upper (resp. lower) half plane of $\lambda$.
The scattering matrix associated to (\ref{eq:Lax-MNLS}) is defined
as
\begin{eqnarray}\label{eq:T_mat}
&&T(t,\lambda)=(\psi(x,t,\lambda))^{-1}\phi(x,t,\lambda) =
\left(\begin{array}{cc} a^+(t,\lambda) & -b^-(t,\lambda)\\
b^+(t,\lambda) & a^-(t,\lambda)\\ \end{array} \right), \nonumber
\\&&
(T(t,\lambda ))^{-1}= \left(\begin{array}{cc} c^-(t,\lambda) &
d^-(t,\lambda)\\ -d^+(t,\lambda) & c^+(t,\lambda)\\ \end{array}
\right),
\end{eqnarray}
where $a^\pm(t,\lambda)$ and $b^\pm(t,\lambda)$ are $2\times 2 $
block matrices. The blocks $a^\pm$, $b^\pm$, $c^\pm$ and $d^\pm$
satisfy a number of relations \cite{GGK05a,GGK05b}; for example
\begin{equation}\label{eq:T-hT}
a^+(\lambda) c^-(\lambda) + b^-(\lambda)d^+(\lambda)=\openone,
\qquad a^+(\lambda) d^-(\lambda) - b^-(\lambda)c^+(\lambda)=0,
\end{equation}
etc. The fundamental analytic solutions (FAS) $\chi^{\pm}
(x,t,\lambda ) $ of $L(\lambda ) $ are analytic functions of
$\lambda $ for $\mbox{Im}\,\lambda \gtrless 0$ and are related to
the Jost solutions by:
\begin{equation}\label{eq:FAS_J}
\chi ^\pm(x,t,\lambda ) = \phi (x,t,\lambda ) S_{J}^{\pm}
(t,\lambda ) = \psi (x,t,\lambda ) T_{J}^{\mp}(t,\lambda ).
\end{equation}
Here $S_{J}^{\pm} $, $T_{J}^{\pm} $ upper- and lower-
block-triangular matrices:
\begin{eqnarray*}\label{eq:S_Jpm}
&&S_J^+(t,\lambda )= \left(\begin{array}{cc} \openone & d^-(t,\lambda)\\
0 & c^+(t,\lambda)\\ \end{array} \right), \quad \hspace{-0.25cm}
S_J^-(t,\lambda )= \left(\begin{array}{cc} c^-(t,\lambda) & 0 \\
-d^+(t,\lambda) & \openone \\ \end{array} \right),\nonumber \\&&
T_J^+(t,\lambda )= \left(\begin{array}{cc} \openone & -b^-(t,
\lambda)\\ 0 & a^-(t,\lambda)\\ \end{array} \right), \quad
\hspace{-0.25cm}
T_J^-(t,\lambda )= \left(\begin{array}{cc} a^+(t,\lambda) & 0 \\
b^+(t,\lambda) & \openone \\ \end{array} \right),
\end{eqnarray*}
satisfying $T_J^\pm(t,\lambda ) \hat{S}_J^\pm (t,\lambda
)=T(t,\lambda ) $ and can be viewed as the factors of a
generalized Gauss decompositions of $T(t,\lambda ) $ \cite{VSG1}.
If $Q(x,t) $ evolves according to (\ref{eq:1}) then the scattering
matrix and its elements satisfy the following linear evolution
equations:
\begin{equation}\label{eq:evol}
i { d b^{\pm} \over  d t } +2\lambda ^2 b^{\pm}(t,\lambda ) =0,
\qquad  i { d a^{\pm} \over  d t }  =0,
\end{equation}
so the block-matrices $a^{\pm}(\lambda)$ can be considered as
generating functionals of the integrals of motion. The fact that
all 4 matrix elements of $a^+(\lambda)$ for $\lambda \in \bbbc_+$
(resp.  of $a^-(\lambda)$ for $\lambda \in \bbbc_-$) generate
integrals of motion reflect the superintegrability of the model
and are due to the degeneracy of the dispersion law of
(\ref{eq:1}).

The system (\ref{eq:1}) can be written in a Hamiltonian form by
introducing the Poisson brackets:
\begin{equation}\label{eq:PB_C2}
\{q_j(x), p_k(y)\} = 2 i \delta _{kj}\delta (x-y), \qquad \{
q_{12}(x), p_{12}(y)\} = i \delta (x-y),
\end{equation}
and the Hamiltonian $H=H_{\rm kin}+H_{\rm int}$:
\begin{eqnarray}\label{eq:H_C2}
H_{\rm kin}&=& \int_{-\infty}^{\infty}dx\, \left({\partial
\Phi_{0} \over
\partial x}{\partial \Phi_{0}^{*} \over
\partial x} +\frac{1}{2}\left({\partial \Phi_{1} \over \partial
x}{\partial \Phi_{1}^{*} \over
\partial x}+{\partial \Phi_{-1} \over \partial
x}{\partial \Phi_{-1}^{*} \over
\partial x}\right)\right),\\
H_{\rm int}=&-& \int_{-\infty}^{\infty}dx\, \left( (|\Phi_{0}|^2
+|\Phi_{1}|^2 )^2+(|\Phi_{0}|^2 +|\Phi_{-1}|^2 )^2 \right)
\nonumber\\&-& \int_{-\infty}^{\infty}dx\,\left(
|\Phi_{0}\Phi_{-1}^{*} +\Phi_{1}\Phi_{0}^{*}|^2 \right) .\nonumber
\end{eqnarray}

As mentioned above, one can use any of the matrix elements of
$a^\pm(\lambda)$ as generating functional of integrals of motion
of our model. Generically such integrals would have non-local
densities and will not be in involution.

The classical $R$-matrix approach \cite{FaTa,ForKu*83} is an
effective method to determine the generating functionals of local
integrals of motion which are in involution. From it there follows
that such integrals are generated by expanding $\ln
m_k^\pm(\lambda)$ over the inverse powers of $\lambda$, see
\cite{VSG1}. Here $m_k^\pm(\lambda)$ are the principal minors of
$T(\lambda)$; in our case \begin{eqnarray} &&m_1^+(\lambda) =
a_{11}^+(\lambda), \qquad m_2^+(\lambda) = \det a^+(\lambda),
\nonumber\\&& m_1^-(\lambda) = a_{22}^-(\lambda), \qquad
m_2^-(\lambda) = \det a^-(\lambda).\end{eqnarray} If we consider
\[ \ln m_k^+(\lambda) = \sum_{s=1}^\infty \lambda^{-k} I_s^{(k)},
\] then one can prove that the densities of $I_s^{(k)}$ are local
in $Q(x,t)$. The fact that \cite{VSG1}:
\[ \{ m_k^\pm (\lambda), m_j^\pm (\mu)\}=0, \qquad \mbox{for}\;
k,j =1,2,\] and for all $\lambda, \mu \in \bbbc_\pm$ allow one to
conclude that $\{I_s^{(k)}, I_p^{(j)}\}=0 $ for all $k,j=1,2$ and
$s,p \geq 1$.

In particular, the Hamiltonian of our model is proportional to
$I_3^{(2)}$, i.e. $H=8iI_3^{(2)}$.


\section{Soliton solutions for the spinor BEC: The $so(5)$ connection}\label{sec:3}

The soliton solutions of the $\mathfrak{sp}(4)$ MNLS (\ref{eq:1})
were derived by using the dressing method \cite{I04}. They can be
considered as particular cases of the soliton solutions of the
generic MNLS eqs., derived through the matrix version of the
Gel'fand-Levitan-Marchenko equation, see
\cite{AblPrinTru*04,IMW04,I04}. Here we extend further these
results and combining the ideas of \cite{I04,VG-DJ} we specify
three types of solitons for the model (\ref{eq:1}).

We start our analysis with the well-known isomorphism between the
algebras $sp(4, \bbbc)$ and $so(5, \bbbc)$ \cite{Helg}. Since the
Lax representation is of pure algebraical nature it is natural to
expect that our model (\ref{eq:1}) can be treated also by an
equivalent Lax operator $L'$ whose potential $Q'(x,t)$ and $J'$
take values in $so(5)$. A consequence of the above-mentioned
isomorphism is that the typical representation of $sp(4)$ used
above is equivalent to the spinor representation of $so(5)$.

So we first remind some of our results in \cite{1,GGK05a,GGK05b},
where we have constructed the fundamental analytic solutions, the
dressing factors, the soliton solutions etc. for a class of Lax
operators (including $L'$), related to the simple Lie algebra
$\mathfrak{g}$, in the typical representation of $\mathfrak{g}$.
So we first have to specify (if necessary) $\mathfrak{g}\simeq
so(5)$ and then reformulate the corresponding results for the
spinor representation of $so(5)$.

The main goal of the dressing method is, starting from a solution
$\chi ^\pm_0(x,t,\lambda ) $ of $L_0(\lambda) $ with potential
$Q_{(0)}(x,t)$ to construct a new singular solution $\chi
^\pm_1(x,t,\lambda )$ with singularities located at prescribed
positions $\lambda _1^\pm $; the reduction ${\bf p} ={\bf q}^\dag
$ used in eq. (\ref{eq:C2.1}) ensures that
$\lambda_1^-=(\lambda_1^+)^*$. The new solutions $\chi
^\pm_1(x,t,\lambda )$ will correspond to a potential
$Q_{(1)}(x,t)$ of $L(\lambda) $ (\ref{eq:Lax-MNLS}) with two
discrete eigenvalues $\lambda _1^\pm $.  It is related to the
regular one by a dressing factor $u(x,\lambda )$
\begin{equation}\label{eq:Dressfactor}
\chi^{\pm}_1(x,t,\lambda)=u(x,\lambda) \chi^{\pm}_0(x,t,\lambda)
u_{-}^{-1}(\lambda ). \qquad u_-(\lambda )=\lim_{x\to -\infty }
u(x,\lambda )
\end{equation}
Note that $u_-(\lambda )$ is a diagonal matrix. The dressing
factor $u(x,\lambda ) $ must satisfy the equation
\begin{equation}\label{eq:u-eq}
i {du  \over dx } + Q_{(1)}(x) u - u Q_{(0)}(x)- \lambda
[J,u(x,\lambda)] =0,
\end{equation}
and the normalization condition $\lim_{\lambda \to\infty }
u(x,\lambda ) =\openone $. Besides $\chi ^\pm_i (x,\lambda ) $,
$i=0,1 $ and $u(x,\lambda ) $ must belong to the corresponding Lie
group $Sp(4,\bbbc) $; in addition $u(x,\lambda ) $ by construction
has poles and/or zeroes at $\lambda_1^\pm$.

The construction of $u(x,\lambda ) $ is based on an appropriate
anzats specifying explicitly the form of its $\lambda
$-dependence.
\begin{eqnarray}\label{eq:rank1}
u(x,\lambda)=&&\openone +\left(c_1(\lambda)-1\right)P_1(x,t)
+\left({1\over c_1(\lambda)}-1\right)\overline{P}_1(x,t),
\nonumber\\&& c_1(\lambda) = {\lambda - \lambda_1^+\over \lambda -
\lambda_1^-},
\end{eqnarray}
where the projectors $P_1(x,t)$ and $\overline{P}_1(x,t)$ are are
of rank 1 and are related by $\overline{P}_1(x)=SP_1^T(x)S^{-1}$.
They must satisfy
$\overline{P}_1(x,t)P_1(x,t)=P_1(x,t)\overline{P}_1(x,t)=0$. By
$S$ we have denoted the special matrix which enters in the
definition of the orthogonal algebra, i.e. $X\in so(5)$ if
$X+SX^TS^{-1}=0$. In the typical representation of $so(5)$ we have
$S=\sum_{k=1}^5 (-1)^{k+1}E_{k,6-k}$ where
$(E_{ij})_{km}=\delta_{ik}\delta_{jm}$. The construction of
$P_1(x,t)$ and $\overline{P}_1(x,t)$ using the `polarization'
vectors is outlined in \cite{GGK05a} and we skip it. The new
potential is obtained from
\[ Q_{(1)}(x,t) - Q_{(0)}(x,t) = (\lambda_1^+ - \lambda_1^-) [ J,
P_1(x,t)- \overline{P}_1(x,t)] .\]

Here we show that the $\lambda$-dependence of $u(x,\lambda)$ may
depend \cite{1} on the choice of the representation of
$so(5,\bbbc)\simeq sp(4,\bbbc)$). For $so(5)$ it was
shown\cite{1,GGK05a,GGK05b,VG-DJ} that there are three types of
solitons:

\begin{itemize}

\item  the first  type of soliton solutions are generated by
dressing factors of the form (\ref{eq:rank1}). For generic choice
of the polarization vectors $P_1(x,t)- \overline{P}_1(x,t) \in
so(5)$.

\item  the second type of soliton solutions are generated
analogously with dressing factor (\ref{eq:rank1}), but due to a
specific choice of the polarization vectors $P_1(x,t)-
\overline{P}_1(x,t) \in so(3)\subset so(5)$.

\item  the third type of soliton solutions are generated again by
(\ref{eq:rank1}) but now the corresponding projectors $P_1(x,t)$
and $\overline{P}_1(x,t)$ have rank 2.

\end{itemize}

Each of these types of soliton solutions have their counterpart
relevant to our model on $sp(4)$.  To the first type of soliton
solutions there correspond dressing factor and potential
$Q_{(1)}(x,t)$ of the form\cite{I04}:
\begin{eqnarray}\label{eq:rank2}
&&\tilde{u}(x,\lambda)=\sqrt{c_1(\lambda)} \pi_1(x) +{1\over
\sqrt{c_1(\lambda)}}\overline{\pi}_1(x), \nonumber\\&&
Q_{(1)}(x,t)- Q_{(0)}(x,t) = {1\over 2} [J, \pi_1(x,t) -
\overline{\pi}_1(x,t)],
\end{eqnarray}
where $\pi_1(x)$ and $\overline{\pi}_1(x)$ are {\it rank 2}
projectors, such that
\begin {eqnarray}\overline{\pi}_1(x)\pi_1(x)=\pi_1(x),\quad
\overline{\pi}_1(x)=0,\qquad
\overline{\pi}_1(x)+\pi_1(x)=\openone.
\end{eqnarray}
This last property ensures the non-degeneracy of $u(x,\lambda)$.
Note that now the dressing factor is not a rational function of
$\lambda$ but for the dressed FAS $\chi(x,\lambda)$  eq.
(\ref{eq:Dressfactor}) we get:
\begin{eqnarray}\label{eq:drs}
\chi^{\pm}_1(x,t,\lambda)=&&\left( \pi_1(x,t) + {1\over
c_1(\lambda)}\overline{\pi}_1(x,t) \right)
\chi^{\pm}_0(x,t,\lambda)\left( \pi_1^- + c_1(\lambda)
\overline{\pi}_1^- \right), \nonumber \\&& \pi_1^-=\lim_{x\to
-\infty} \pi_1(x,t),
\end{eqnarray}
i.e., the fractional powers of $c_1(\lambda)$ disappear.

The second type of solitons with rank 2 projector $P_1(x)$ after
recalculating to the spinor representation formally keeps the same
form (\ref{eq:rank1}) with $P_1(x)$ replaced by $A_1(x)$ which has
rank 1 but generically is not a projector, see \cite{I04}.

The third type of solitons is similar to the second one but with
additional constraints on the factor $A_1(x)$ so that $A_1(x) -
\overline{A}_1(x)\in sp(2)\subset sp(4)$.

Consider the purely solitonic case when $Q_{(0)}=0$. From now on
we introduce the following notations $\lambda _1^{\pm} = \mu _1
\pm i \nu_1$ and
\begin{eqnarray}\label{AB}
&&A = -2 i ((\lambda_1^+ )^2 -
 (\lambda_1^{-} )^2)t-i (\lambda_1^+ - \lambda_1^- )x,\nonumber
 \\&&
 B  =  -2 ((\lambda_1^+ )^2 + (\lambda_1^{-})^2)t-
(\lambda_1^+ +\lambda_1^- )x.
\end{eqnarray}
Here $A(x,t)$ and $B(x,t)$ are $x$ and $t$ dependent real valued
functions. Making use of the explicit form of the projectors
$P_{\pm 1}(x) $ valid for the typical representations of ${\bf
B}_2 $ we obtain\cite{GGK05a}:
\begin{eqnarray}\label{eq:26.11}
\Phi_{(1)}(x,t)= \frac{4(\lambda _1^+ - \lambda _1^-)}{\langle m|n
\rangle } \left( n_{0,1}m_{0,2} e^{A}+ n_{0,\bar{2}} m_{0,\bar{1}}
e^{-A}\right) e^{i B}\\
\Phi_{(0)}(x,t)=i \frac{2 \sqrt{2} (\lambda _1^+ - \lambda
_1^-)}{\langle m|n \rangle } \left( n_{0,1}m_{0,3} e^{A}-
n_{0,\bar{3}} m_{0,\bar{1}} e^{-A} \right)e^{i B} \\
\Phi_{(-1)}(x,t)= -\frac{4(\lambda _1^+ - \lambda _1^-)}{\langle
m|n \rangle } \left(  n_{0,1}m_{0,\bar{2}} e^{A}+ n_{0,2}
m_{0,\bar{1}} e^{-A} \right)e^{i B}
\end{eqnarray}
where the denominator in the above formula is given by:
\begin{eqnarray}\label{<m|n>}
 \langle m|n \rangle &=& m_{0,1}n_{0,1}
(e^{2A})  + m_{0,\bar{1}}n_{0,\bar{1}}(e^{-2A})\nonumber\\
 &+& m_{0,2}n_{0,2} + m_{0,\bar{2}}n_{0,\bar{2}}+ m_{0,3}n_{0,3} .
\end{eqnarray}
and $m_{0,k}$, $n_{0,k}$ are the components of the polarization
vectors.

Choosing appropriately the elements of the polarization vectors
$|n_0 \rangle$ and $|m_0 \rangle$, one can show that the
conjecture that the Zakharov-Shabat dressing procedure and the
Gel'fand-Levitan Marchenko formalism lead to comparable soliton
solutions is true. It is not a problem to multiply the
polarization vectors $|n_0 \rangle$ and $|m_0 \rangle$ by an
appropriate scalar and thus to adjust the two solutions. Such a
multiplication easily goes through the whole scheme outlined
above. The involution $Q_{(1)}^{\dag}=Q_{(1)}$ that the potential
of the Lax operator (\ref{eq:Lax-MNLS}) associated with the system
(\ref{eq:1}) is subject to results in the following relations
between the elements of the ''polarization'' vectors $|n_0\rangle
$ and $\langle m_0| $, namely $ n_{0,k}=m_{0,k}^{\ast}$. Utilizing
the above and a proper change of field components, we can relate
the solution
\begin{equation}\label{1a-sol}
 {\bf q}(x,t)
 =4\nu_1\; \frac{ C^{\dagger}e^{A}+
 \sigma_{2} C^{t}\sigma_{2} \det\{ C^{\dagger} \}
e^{-A}}{ e^{2A} + W +
 |\det\{C\}|^2
e^{-2A}}\;e^{i B},
 \end{equation}
where $W=(2|c_{12}|^2+|c_{1}|^2+|c_{2}|^2)$ and the
''polarization'' matrix can be cast into the form
\begin{equation}\label{C}
    C= \left(
\begin{array}{cccc}
c_{12} & c_{1} \\
 c_{2} &  -c_{12} \\
 \end{array}
 \right)
\end{equation} In the special
case when $W=1$ and $\det\{C\}=0$ we obtain
\begin{equation}\label{1aa-sol}
 {\bf q}(x,t)=\frac{2\nu_1 e^{i B}}{ \cosh{A}}\; C^{\dagger}
\end{equation}

Thus we confirm the result obtained in\cite{IMW04}, aquaired with
the help of GLM formalism and the solution (\ref{eq:26.11}),
derived within the generalized Zakharov-Shabat dressing procedure,
provided we make sure that the extra condition on  the vector
$|m\rangle $:
\begin{eqnarray}\label{eq:nSn}
 -2 m_{0,1} m_{0,\bar{1}} + 2 m_{0,2} m_{0,\bar{2}} =
\left(m_{0,3}\right)^2 ,
\end{eqnarray}
and analogous one for $|n\rangle $  holds true. Setting
\begin{eqnarray*}
&&
  m_{0,1} = 1, \quad
  m_{0,\bar{1}} = - (c_{12}^{\ast})^2 - c_1^{\ast} c_2^{\ast}
  ,\nonumber \\&&
  m_{0,2} = i c_1^{\ast},  \quad  m_{0,\bar{2}} = i c_2^{\ast}, \quad
  m_{0,3} = m_{0,\bar{3}}  =  -\sqrt{2} c_{12}^{\ast}
\end{eqnarray*}
we establish the equivalence between the two solutions.

\section{Conclusions}\label{sec:Concl}

We have derived the soliton solutions of the three-component
system of NLS type on the symmetric space ${\rm Sp(4)}/{\rm U(2)}$
which is related to spinor  Bose-Einstein condensate model (with
$F=1$). Furthermore, we have described briefly the Hamiltonian
properties of the model and the  integrals of motion. Using the
classical $r$-matrix approach, we showed that the integrals of
motion, that belong to the principal series are in involition.

The reduction of the multi-component nonlinear Schr\"{o}dinger
(NLS) equations on symmetric space ${\bf C.I}\simeq {\rm
Sp(2p)}/{\rm U(p)}$ for $p=2 $ is related to spinor model of
Bose-Einstein condensate. Other interesting reductions of MNLS
type equations were reported in \cite{GGK05a} and a systematic
study of the problem is on the way. Recently the authors of
\cite{djj08} develop a perturbation theory for bright solitons of
the $F = 1$ integrable spinor BEC model. Both rank-one and
rank-two soliton solutions are obtained using Riemann-Hilbert
method  and are compared with known results.

\subsection*{Acknowledgments}\label{sec:Ack}

This work has been supported also by the National Science
Foundation of Bulgaria, contract No. F-1410.

\label{lastpage}


\begin{thebibliography}{[MT1]}

\small
\bibitem{IMW04} \textsc{Ieda J}, \textsc{Miyakawa T} and \textsc{Wadati M},  Exact
Analysis of Soliton Dynamics in Spinor Bose-Einstein Condensates,
{\it Phys. Rev Lett.} {\bf 93} (2004), 194102.

\bibitem{AblPrinTru*04} \textsc{Ablowitz M J}, \textsc{Prinari B}, and  \textsc{Trubatch A D}  Discrete and
continuous nonlinear Schr\"odinger systems, Cambridge Univ. Press,
Cambridge, 2004.

\bibitem{I04} \textsc{Ivanov R I}, On the Dressing Method for the
Generalized Zakharov-Shabat System, {\it Nucl. Phys.} {\bf B 694},
(2004) 509--524.

\bibitem{ForKu*83} \textsc{Fordy A P} and \textsc{Kulish P P},  Nonlinear Schrodinger
Equations and Simple Lie Algebras, {\it Commun. Math. Phys.} {\bf
89} (1983) 427--443.

\bibitem{LLMML05} \textsc{Li L}, \textsc{Li Z}, \textsc{Malomed B A}, \textsc{Mihalache D}  and
\textsc{Liu W M},  Exact Soliton Solutions and Nonlinear
Modulation Instability in Spinor Bose-Einstein Condensates, {\it
Phys. Rev. A} {\bf 72}, (2005) 033611.

\bibitem{FaTa} \textsc{Faddeev L D} and \textsc{Takhtadjan L A},  Hamiltonian
Approach in the Theory of Solitons, Springer Verlag, Berlin, 1987.

\bibitem{ZMNP} \textsc{Zakharov V E}, \textsc{Manakov S V}, \textsc{Novikov S P}  and
\textsc{Pitaevskii L I} ,  Theory of Solitons. The Inverse
Scattering Method, Plenum Press (Consultant Bureau), N.Y., 1984.

\bibitem{Helg} \textsc{Helgasson S},
Differential Geometry, Lie Groups and Symmetric Spaces, Graduate
studies in Mathematics, vol.34, AMS, Providence, Rhod Island 2001.

\bibitem{Za*Mi} \textsc{Zakharov V E} and \textsc{Mikhailov A V}, On the
Integrability of Classical Spinor Models in Two--dimensional
Space--time, {\it Commun. Math. Phys.} {\bf 74} (1980) 21--40.

\bibitem{1} \textsc{Gerdjikov V S}, \textsc{Grahovski G G}, \textsc{Ivanov R I} and \textsc{Kostov N A},
$N $-wave Interactions Related to Simple Lie Algebras. $\bbbz_2$-
reductions and Soliton Solutions, {\it Inv. Problems} {\bf 17}
(2001) 999--1015.

\bibitem{GGK05a} \textsc{Gerdjikov V S}, \textsc{Grahovski G G}  and \textsc{Kostov N A}, On the
multi-component NLS type equations on symmetric spaces and their
reductions, {\it Theor. Math. Phys.} {\bf 144} (2005), No.2,
1147--1156.

\bibitem{GGK05b} \textsc{Grahovski G G}, \textsc{Gerdjikov V S}, \textsc{Kostov N A},
and  \textsc{Atanasov V A}, New Integrable Multi-component NLS
type Equations on Symmetric Spaces: $Z_4$ and $Z_6$ reductions, In
{\it ``Geometry, Integrability and Quantization VII''}, Eds: I.
Mladenov and M. De Leon, Softex, Sofia (2006), pp. 154--175;
(E-print: {\tt nlin.SI/0603066}).

\bibitem{VSG1} \textsc{Gerdjikov V S},
 The Zakharov--Shabat dressing method amd the
 representation theory of the semisimple Lie algebras,
     {\it Phys. Lett. A} {\bf 126A,} n.~3, 184--188, (1987).

\bibitem{VSG2}  \textsc{Gerdjikov V S}.
 Basic Aspects of Soliton Theory. In: Eds.: I. M. Mladenov, A.
C. Hirshfeld. "Geometry, Integrability and Quantization", Softex,
Sofia 2005. {\bf nlin.SI/0604004}, 78--125

\bibitem{VG-DJ} \textsc{Gerdjikov V S},  \textsc{Kaup D J},
 How many types of soliton solutions do we know? In: Eds. Ivailo
Mladenov, Manuel de Leon.  "Geometry, Integrability and
Quantization", Softex, Sofia, 2006, 11--34.

\bibitem{djj08} \textsc{Doktorov E V}, \textsc{Wang J} and
\textsc{Yang J}, Perturbation theory for bright spinor Bose
Einstein condensate solitons,
 {\it in press, private comminication }, 2008.



\end{thebibliography}
\end{document}